*Correspondence and requests for materials should be addressed to
S. K. (stephan.krohns@physik.uni-augsburg.de)

# Magnetoelectric effects in the skyrmion host material $Cu_2OSeO_3$

E. Ruff[1], P. Lunkenheimer[1], A. Loidl[1], H. Berger[2] & S. Krohns[1,*]

[1]Experimental Physics V, Center for Electronic Correlations and Magnetism, University of Augsburg, 86135 Augsburg, Germany
[2]Institute of Physics of Complex Matter, EPFL, CH-1015 Lausanne, Switzerland

**Insulating helimagnetic $Cu_2OSeO_3$ shows sizeable magnetoelectric effects in its skyrmion phase. Using magnetization measurements, magneto-current analysis and dielectric spectroscopy, we provide a thorough investigation of magnetoelectric coupling, polarization and dielectric constants of the ordered magnetic and polar phases of single-crystalline $Cu_2OSeO_3$ in external magnetic fields up to 150 mT and at temperatures below 60 K. From these measurements we construct a detailed phase diagram. Especially, the skyrmion phase and the metamagnetic transition of helical to conical spin order are characterized in detail. Finally we address the question if there is any signature of polar order that can be switched by an external electric field, which would imply multiferroic behaviour of $Cu_2OSeO_3$.**

Materials characterized by the simultaneous existence of magnetic and polar order belong to the group of multiferroics[1,2] and allow for detailed studies of magnetoelectric couplings, which will be of prime importance for future technological applications. Materials with non-collinear spin structures are promising candidates for multiferroic behaviour as, e.g., demonstrated for $TbMnO_3$[3] or $LiCuVO_4$[4,5,6]. Insulating $Cu_2OSeO_3$, exhibiting helical spin structures, also shows magnetoelectric coupling below 60 K[7,8]. However, $Cu_2OSeO_3$ was also reported to display the formation of a skyrmion crystal (SkX)[9,10,11]. Skyrmions[12] are stable vortex-like topological spin structures, which can be treated as particles, with dimensions in the nanometer range. In metallic systems like MnSi[13] or $Fe_{1-x}Co_xSi$[14], these particle-like vortices are electrically controllable making them interesting for magnetic storage applications[15]. In insulating systems as in $Cu_2OSeO_3$, two neighbouring spins $\mathbf{S_i}$ and $\mathbf{S_j}$ of the chiral spin structure possibly can induce electrically ordered phases via the inverse Dyzaloshinskii-Moriya (DM) interaction which depends on $(\mathbf{S_i} \times \mathbf{S_j})$[16,17,18]. This is the leading interaction in spin-driven multiferroics and allows for switching of the electric polarization by external magnetic fields like, e.g., in $LiCuVO_4$[4,5,6]. For skyrmion crystals of chiral magnets there are two further possible mechanisms for the coupling of magnetic and electric order, namely an $(\mathbf{S_i} \cdot \mathbf{S_j})$ Heisenberg-like exchange interaction and the *d-p* hybridization[13,14,19,20].

Insulating $Cu_2OSeO_3$ crystallizes in the non-centrosymmetric $P2_13$ space group[21] and exhibits complex magnetic exchange coupling of $Cu^{2+}$ ions, which are coordinated within square pyramidal and trigonal bipyramidal $CuO_5$ in the ratio of 3:1 within the unit cell[9,22]. Below the magnetic ordering temperature of about $T_c$ = 59 K, this system forms helical, conical, ferrimagnetic and skyrmion crystal phases in an applied external magnetic field and thus allows, due to possible ferroelectricity based on non-centrosymetric space group, the analysis of the impact of magnetically ordered states on polar phases. For this system it was demonstrated that $(\mathbf{S_i} \cdot \mathbf{r_{ij}})^2$-like exchange interaction, i.e., *d-p* hybridization, where $r_{ij}$ is the bonding vector between $Cu^{2+}$ and $O^{2-}$ ions, is the most appropriate magnetoelectric mechanism which however, does not lead to an electric control of the polarization like in conventional ferroelectrics[19,20,23].

In the present work, via magnetization and magneto-current measurements we thoroughly investigate the magnetoelectric phases of the skyrmion-crystal host $Cu_2OSeO_3$ and provide a detailed magnetic and electric phase diagram for two bulk samples. In addition, results of dielectric spectroscopy in applied magnetic fields are reported. We do not find conclusive indications of a multiferroic skyrmion phase in the sense of switchable polar order coexisting with spin order, but we document that polar order is only induced by external magnetic fields. However, there are strong magnetoelectric effects in the helical and conical phases, but also in the skyrmion crystal. Furthermore, we test the possibility of electric control of the polarization applying electric fields in crystallographic ⟨111⟩ direction during sample cooling in zero magnetic field into the helical and with magnetic field into the SkX phase, respectively.

**Magnetization**

Previously, detailed and systematic magnetizations measurements in applied magnetic fields up to 150 mT at various temperatures between 4 and 60 K were used to identify the magnetic phases of the skyrmion-host crystal $Cu_2OSeO_3$[9,22]. Below the magnetic ordering temperature of about $T_c = 59$ K and on increasing external magnetic fields, the system exhibits helical, conical and field-induced spin-collinear (ferrimagnetic) phases[9,22,24]. The formation of the skyrmion crystal phase appears in a relatively restricted temperature and magnetic field range, as an additional phase between helical and field-induced collinear spin order.

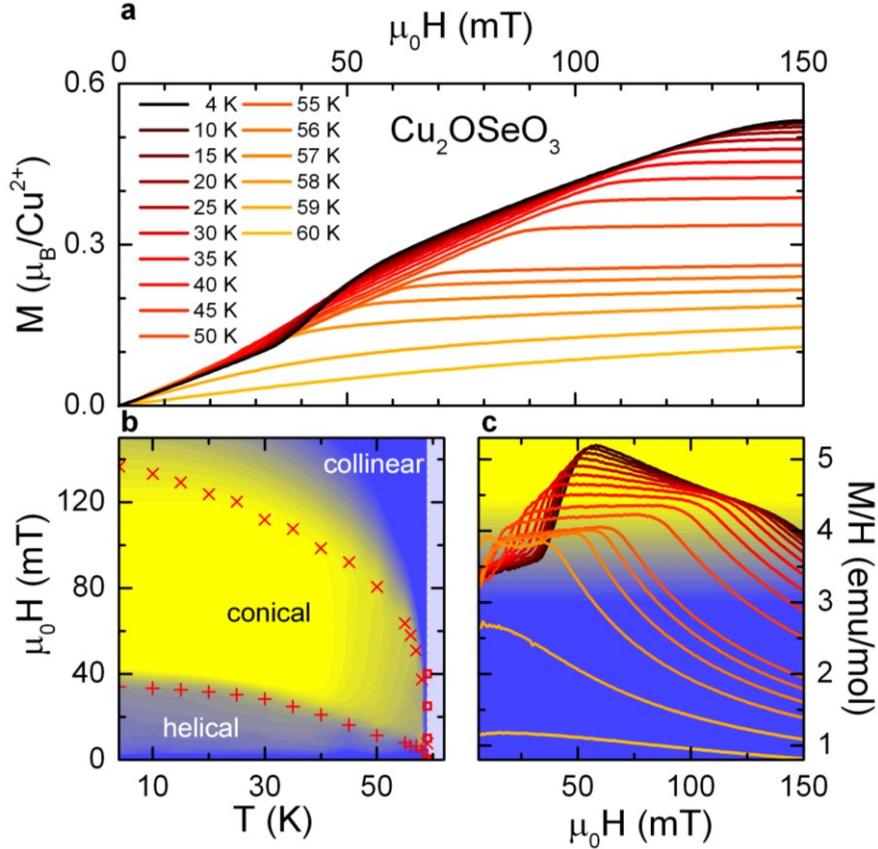

FIG. 1. Magnetic properties of $Cu_2OSeO_3$ for applied magnetic fields H ∥ ⟨111⟩. (a) Magnetic-field dependence of the magnetization M for various temperatures. (b) Plusses and crosses indicate the anomalies showing up in (c) (see text for details). They correspond to metamagnetic transitions from helical to conical (plusses) and from conical to collinear ferrimagnetic spin order (crosses). Superimposed to this, (b) also provides a colour-coded contour plot of M/H versus temperature and external magnetic field. The transition into the paramagnetic phase is marked by squares and the paramagnetic regime is colored in light blue. The colour code for M/H is shown as background colour of frame (c). (c) Lines: Susceptibility M/H versus applied magnetic field for various temperatures as indicated in frame (a).

To identify the magnetic phase boundaries in $Cu_2OSeO_3$, Figure 1 shows the magnetization M versus applied magnetic field (H ∥ ⟨111⟩) for various temperatures (a), a H,T-phase diagram superimposed with a colour-coded plot of M/H versus external magnetic fields from 0 to 135 mT and temperatures ranging from 4 K to 62 K (b) and the susceptibility M/H versus applied magnetic field (c). The changes of slope of the magnetization M vs. applied magnetic field H, as shown in Fig. 1(a), clearly point towards successive metamagnetic transitions. At lower temperatures, e.g., documented for the magnetization at 4 K, the upward curvature at 35 mT indicates the transition from helical to conical spin order with a residual ferromagnetic moment. If the applied magnetic fields are sufficiently strong (e.g., close to 130 mT for the magnetization curve at 15 K), the magnetization saturates at about 0.5 $\mu_B/Cu^{2+}$ and thus indicates the onset of the field-induced collinear ferrimagnetic phase.

To determine these transitions with higher precision, Fig. 1(c) illustrates M/H versus H. A first inspection shows two dominating cusp-like anomalies, one shifting from 15 to 50 mT, the other from 40 to 130 mT, roughly marking the boundaries of the conical spin phase with plateau-like maximal values of M/H at a given temperature. However, the onset of the phase transition from the helical to the conical phase sets in at somewhat lower fields marked by the abrupt increase of M/H, shifting from very low fields to 35 mT at 4 K. This transition from the helical to the conical phase is indicated by plusses in Fig. 1(b). The knee-like decrease of M/H shifting from 40 mT at 58 K to approximately 130 mT at the lowest temperatures is the fingerprint of the transition from the conical to the field-induced ferrimagnetic phase and is indicated by crosses in Figure 1b. The resulting phase boundaries are in perfect agreement with literature[9,22]. Beside the two significant transitions as documented in Figure 1, a closer analysis of the magnetization at 57 K reveals an additional double-peak structure, which indicates the SkX phase, studied in more detail in the following.

From literature[9,22,23], it is known that the SkX phase in $Cu_2OSeO_3$ appears only in a restricted range of temperatures (in the vicinity of $T_c$) and external magnetic fields (around 20 mT). In Figure 2 we show a thorough analysis of dM/dH versus applied magnetic fields H ∥ ⟨111⟩ for temperatures close to $T_c$.), A detailed H,T-phase diagram in the region of the SkX phase at temperature from 55.5 K to 59 K superimposed with a colour-coded plot of dM/dH is provided in Fig. 2(a). The derivative of the magnetization dM/dH [Fig. 2(c)] reveals two distinct anomalies in the temperature regime from 56 K to 58 K. Following refs. 22 and 25, these local maxima in the dM/dH vs. H curves define the upper and lower critical magnetic fields of the SkX phase. The stars in Fig. 2(a) mark these boundaries of the SkX phase as determined in Fig. 2(b). The metamagnetic transitions as shown in Figure 1 are indicated by plusses and crosses. Our detailed analysis of the magnetic order of single crystalline $Cu_2OSeO_3$ nicely confirms the helical, conical and especially the SkX phases as determined previously[9,22].

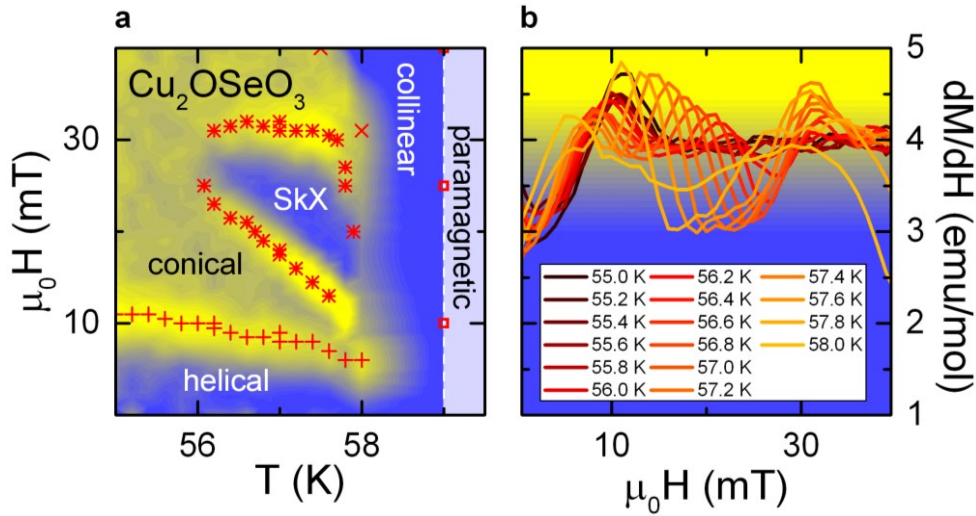

FIG. 2. Magnetic properties of $Cu_2OSeO_3$ in the vicinity of $T_c$ for applied magnetic fields H ∥ ⟨111⟩. (a) H,T-phase diagram in the vicinity of the magnetic phase-transition temperature. Plusses and crosses define the metamagnetic transitions as shown in Fig. 1(b). The stars denote the location of the maxima in dM/dH, which indicate the upper and lower critical magnetic fields of the SkX-phase. Superimposed to this phase diagram, (b) also includes a colour-coded contour plot of the values of the derivative of the susceptibility dM/dH. The colour code is defined by the background colours in frame (b). (b) Lines: dM/dH versus applied magnetic field for temperatures between 55 and 58 K.

**Polarization**

There are at least three possible coupling mechanisms of polarization P and spin order[16,26,27] for spin driven ferroelectrics. However, in insulating $Cu_2OSeO_3$ the d-p hybridization model[26,27] is favoured. To gain a more detailed insight, we performed a thorough analysis of the polarization P as function of external magnetic fields H for temperatures close to $T_c$, specifically focusing on the skyrmion phase. Figure 3a shows the polarization P(H) as derived from magnetocurrent measurements for temperatures in a narrow temperature range between 56 and 58 K just below $T_c$. As known from pyrocurrent measurements (not shown), for all temperatures the polarization is zero for zero external magnetic fields, indicating the absence of spontaneous spin-induced polarization, at least on a macroscopic scale. This probably results from cancellation of the polarization in the multiple q structure of the helical phase. When examining the field dependence at 56 K, on increasing H the polarization first becomes negative, reaching a minimum of -0.5 µC/m² at about 20 mT. At higher fields, P(H) increases again displaying a parabolic behaviour within the conical phase. Above about 70 mT, P(H) saturates within the collinear ferrimagnetic phase at values of about 1.4 µC/m². As documented by magnetization measurements, at this temperature and for H = 0 mT the helical spin structure prevails, leading to a multi q-domain structure[19], while for H > 70 mT the collinear ferrimagnetic spin state is established (cf. Figures 1b and 2a). According to the d-p hybridization model[26,27] ferroelectric polarization is induced along the crystallographic ⟨111⟩ direction.

Following the model of Belesi et al.[23] a flat, non-conical helix induces ferroelectric polarization pointing along the propagation vector $q$ of the helix. Hence, the overall net polarization is averaged to zero due to the four different $q$-directions of the spin helices in the multidomain state. The net polarization in the intermediate magnetic regime is driven by the emerging conical spin helix (H > 10 mT) and the competing polarizations of the multi $q$-domain helices in the helical phase (H < 10 mT). Increasing the external magnetic field leads to both, a reorientation process of the helices and to the formation of conical spin spirals. At the metamagnetic transition from helical to conical spin structure, first a negative polarization – antiparallel to $\langle 111 \rangle$ – appears. For 56 K it reaches a minimum of about -0.5 µC/m² at 20 mT, in perfect agreement with theoretical predictions[23]. At magnetic fields H ∥ $\langle 111 \rangle$ and higher than 20 mT, within the conical phase, out of plane contributions to the polarization emerge. These are parallel to the crystallographic $\langle 111 \rangle$ direction and increase with increasing external magnetic field displaying a quadratic magnetoelectric coupling. The dashed line in Fig. 3(a) represents a simulation of $P(H) \sim M(H)^2$ for the 56 K curve, $P_{coni,56K}(H)$. This spin-driven polarization in the conical phase is accompanied by a transition of the multidomain into a single $q$-domain state. Thus, negative polarization arises due to reorientation of the helices breaking the symmetry between the four $q$-directions and at H > 20mT a positive polarization induced by increasing conical tilting of spins starts to prevail. Finally, at magnetic fields (H > 70 mT) the polarization saturates at values close to 1.4 µC/m², which denotes the second metamagnetic transition from the conical into the ferrimagnetic collinear phase [cf. Figs. 1(b) and 3(b)].

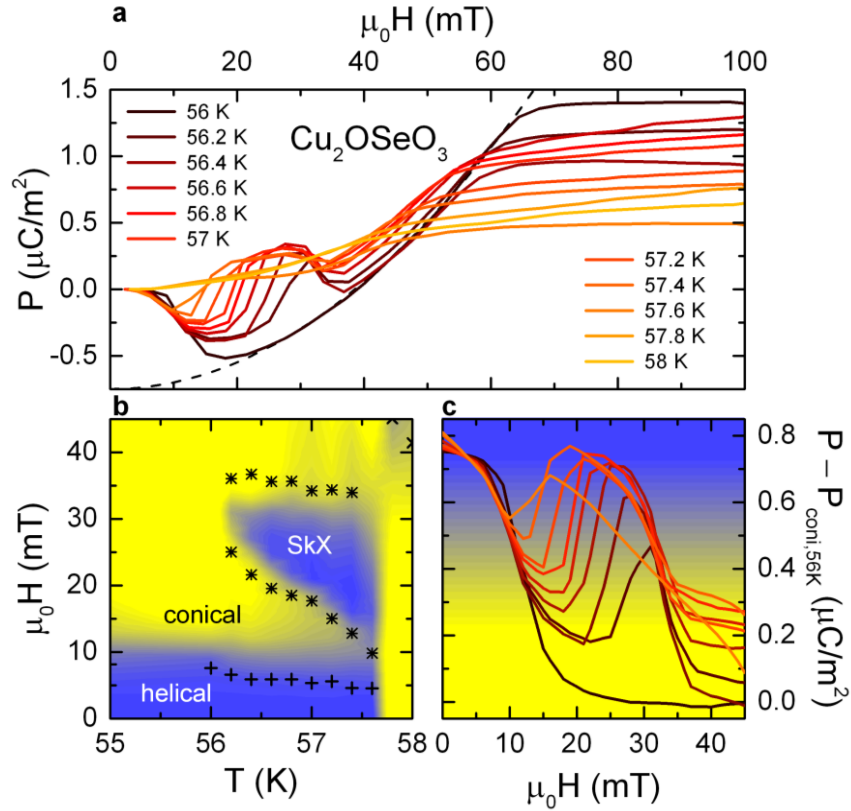

FIG. 3. (a) Magnetic-field dependent polarization P with P and H ∥ $\langle 111 \rangle$ of $Cu_2OSeO_3$ for various temperatures between 56 and 58 K. The dashed line represents a fit $P_{coni,56K}(H)$ of the polarization in the conical phase displaying a quadratic magnetoelectric coupling. (b) H,T-phase diagram. The stars in (b) indicate the phase boundaries of the SkX phase. The plusses document the change in the polarization at the magnetic transitions from helical to conical spin order. Superimposed to the phase diagram, frame (b) also shows a colour-coded plot of the polar contributions arising within the helical and SkX phase (see text for details) in vicinity of $T_c$. The corresponding colour code is indicated as background colour in frame (c). (c) Lines: Magnetic-field dependent net polarization of the helical and SkX phases (normalized by substracting $P_{coni,56K}(H)$) for various temperatures between 56 and 58 K.

Now, we have a closer look to the field dependence of the ferroelectric polarization in the skyrmion crystal phase, which emerges at magnetic fields from 10 to 30 mT and in the temperature range from 56.2 to 58 K. The polarization for these temperatures [Fig. 3(a)] shows significant deviations from the smooth parabolic-like behaviour of P(H) at 56 K that mainly results from the transition of the helical into the conical spin state. In the SkX phase a clear double-well profile of the polarization is detected. The deviations from a purely parabolic behaviour determine the phase boundaries of the SkX state, marked with stars in the magnetic phase diagram of Fig. 3(b). As outlined by Seki et al.[9], the SkX phase is an "intervening" state between helical and spin-collinear phase, which exhibits reduced symmetry. In this phase, only polar contributions parallel to the applied magnetic field persist. In our framework, this leads to an intermediate peak in P(H) in the ⟨111⟩ direction for the SkX state.

The magnetic-field dependent polarization $P_i(H_j)$ determined via magnetocurrent measurements points towards significant magnetoelectric effects, which can be parameterized in the form: $P_i = \alpha_{ij} H_j + \beta_{ijk}/2\, H_j H_k + \ldots$[1]. Here $\alpha_{ij}$ is the linear magnetoelectric coupling term and $\beta_{ijk}$ the quadratic magnetoelectric coefficient. Following the *d-p* hybridization model as well as guided by the parabolic magnetic-field dependence of the polarization, it seems that the magnetic-field dependence of the ferroelectric polarization along the crystallographic ⟨111⟩ direction within the conical phase is given by $P_{[111]} = P_0 + \beta M^2$. Polar contributions of helical and SkX phases lead to a significant deviation from quadratic magnetoelectric coupling, which probably results in an enhanced magnetoelectric coefficient. In Figure 3c the magnetic field dependence of P(H) substracted by the pure polar contributions of the conical phase at 56 K, i.e., $P_{coni,56K}(H)$, is shown for temperatures and magnetic fields around the SkX state. The net polarization reaches values of the order of 0.8 μC/m² in the helical phase. In this regime the polarization is antiparallel to the external magnetic field and compensates the polarization of the conical phase, thus the overall polarization is zero, as determined by pyrocurrent measurements. On further increasing magnetic fields the SkX phase shows up in Fig. 3(c) as a well pronounced peak leading to net polarization of the order of 0.5 μC/m². At temperatures below 56.2 K the SkX phase vanishes and the transition of the polarization from helical to conical ordered regimes remains. Hence, this behaviour is best documented in the measurement at 56 K. For magnetic fields $\mu_0 H > 20$ mT the net polarisation reaches zero and indicates the quadratic magnetoelectric behaviour, leading to a magnetoelectric coupling coefficient of $\beta \sim 0.34 \times 10^{-6}$ μC mol²/m²emu². To summarize the findings documented in Fig. 3(c), we found magnetoelectric coupling in the conical phase and a positive coupling in the SkX phase. It is interesting that in ab-initio calculations, Spaldin and co-workers[28] have found strongly enhanced magnetoelectricity in magnetic vortices, in this case, however, in kagome lattices and only assuming linear magnetoelectric coupling.

**Dielectric spectroscopy**

Dielectric spectroscopy in external magnetic fields is an appropriate technique to identify the polar ground state of materials and to quantify magnetoelectric coupling[2,3,4,6,29,30,31]. Temperature dependences of the dielectric constants and, partly, magnetocapacitance experiments on $Cu_2OSeO_3$ have been reported by Bos et al.[7], Miller et al.[8] and Belesi et al.[23]. Low-temperature dielectric constants between 7 and 15 were determined in these investigations and shown to exhibit a small anomaly when passing the magnetic phase transition. Figure 4 shows the temperature-dependent dielectric constant measured at $\nu = 100$ kHz for various magnetic fields between 0 and 70 mT. The dielectric constant is $\varepsilon' \sim 12.5$ at low temperatures and it slightly increases by about 0.1% at the magnetic phase transition. It is almost independent of magnetic field and only weakly dependent on frequency (inset of Figure 4). The small increase of the dielectric constant in passing the magnetic phase transition probably is due to magnetostriction effects[32]. Due to fluctuation effects, the transition is significantly smeared out in a broad temperature regime between 30 and 80 K. At temperatures below 20 K the dielectric constant reaches saturation for all magnetic fields.

In most cases, even in spin-driven improper ferroelectrics like $TbMnO_3$[3], ferroelectric transitions show up as peak in the temperature dependence of the dielectric constant. However, in $Cu_2OSeO_3$ weak magnetocapacitive effects but no indications of a ferroelectric transition are observed. For zero external magnetic fields, this observation agrees with the absence of a polar state in $Cu_2OSeO_3$ discussed above [cf. Fig. 3(a)]. As already mentioned, it has been argued that in the multiple *q* state the polarization remains zero due to cancellation effects of P averaged over all domains[19,23]. However, Figure 4 documents that even at non-zero magnetic fields, there is no significant signature of ferroelectricity (and thus of multiferroic ordering) in $\varepsilon'(T)$ of $Cu_2OSeO_3$. A close inspection of the magnetic-field dependent dielectric properties reveals a slightly enhanced dielectric constant $\varepsilon'$ in the vicinity of $T_c$ as measured at 25 mT. Exactly at this field, the SkX phase is crossed close to 60 K and the slight enhancement of the dielectric constant might result from polar contributions characteristic of the SkX phase. However, the observed enhancement is only slightly beyond experimental uncertainty.

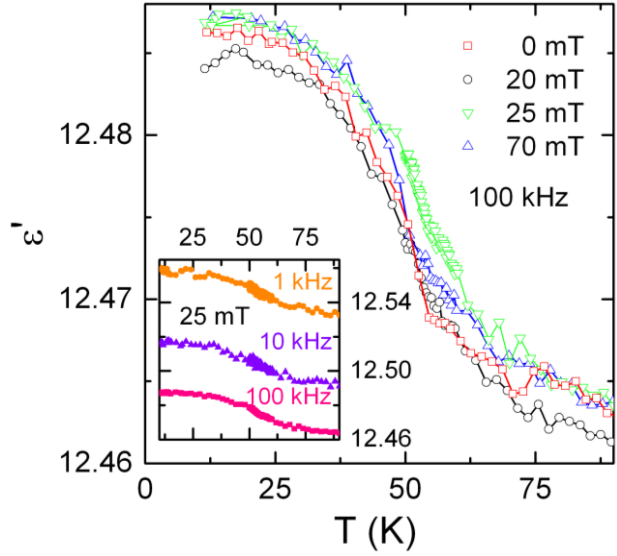

FIG. 4. Temperature-dependent dielectric constant of $Cu_2OSeO_3$. $\varepsilon'(T)$ is shown for different magnetic fields as measured at 100 kHz, with E and H ∥ ⟨111⟩. The inset shows the temperature dependence of $\varepsilon'$ in an external magnetic field of $\mu_0 H$ = 25 mT for different frequencies.

**Phase diagram and poling measurements**

Figure 5a illustrates the H,T-phase diagram of $Cu_2OSeO_3$ for temperatures below $T_c$, comprising results from temperature- and magnetic-field-dependent magnetic susceptibility and magnetoelectric-coefficient measurements. The magnetic phase boundaries from helical to conical and from conical to field-induced collinear spin order are in good agreement with previous published phase diagrams of bulk $Cu_2OSeO_3$ samples [9,22,24,33]. The temperature and magnetic-field dependence of the emerging SkX-phase in the vicinity of $T_c$ [Fig. 5(b)] is also well verified by our measurements.

Notably, the electrical control of the particle-like vortices in this material is still an interesting question[9,11]. It seems possible via the magnetoelectric coupling within the SkX phase[9]. Following suggestions of ref. 11, the electrical control of magnetic vortices within the SkX phase should be achievable for electric and magnetic fields E ∥ H ∥ ⟨111⟩. The *d-p* hybridization seems to be the most reasonable mechanism for magnetoelectric coupling, but in this model ferroelectricity is only induced by external magnetic fields and $Cu_2OSeO_3$ is not a canonical multiferroic with spontaneous magnetic and ferroelectric polarization.

The definition of ferroelectricity implies a unique polar axis and the switchability of the polarization by an electrical field[34]. To search for a possible switching behaviour of the polarization in $Cu_2OSeO_3$, we have carried out magnetocurrent measurements applying positive and negative electrical prepoling fields. These poling experiments were performed on a second sample with significantly reduced thickness ($d_{sample1}$ = 2.1 mm *vs.* $d_{sample2}$ = 0.35 mm) which enabled reaching high electrical fields of $E_0$ = ± 330 kV/m. Figure 5c shows the polarization determined from these magnetocurrent measurements performed at 57 K. The blue lines show the results of two measurement runs that were carried out after the sample was cooled from above $T_c$ to 57 K with electrical dc fields of + 330 kV/m (blue solid line) and - 330 kV/m (blue dashed line) applied along ⟨111⟩. During prepoling no magnetic field was applied, i.e., the measurements started in the helical phase [see blue arrow in Fig. 5(b)]. The polarization curves determined from these magnetocurrent measurements qualitatively resemble the P(H) results shown in Fig. 3(a). Switching the direction of the prepoling field only results in minor changes in P(H), which also becomes obvious in the inset of Fig. 5(c) showing the difference in polarization $\Delta P = P_{+E}(H) - P_{-E}(H)$. Such behaviour clearly is not expected for a canonical ferroelectric and the polarization in the helical phase of $Cu_2OSeO_3$ obviously could not be switched by the electrical field. This finding is in good accord with the formation of multidomain helices with zero net polarization, discussed above, which impedes the establishment of a preferred orientation of the polarization by an external electrical poling field.

Interestingly, if the SkX phase is directly accessed as the starting phase for the magnetocurrent measurement and prepoling is done within this phase, there still is no canonical ferroelectric switching behaviour of the polarization. The corresponding experiment [cf. red arrow in Fig. 5(b)] was performed by applying an external magnetic field of $\mu_0 H$ = 20 mT during the preceding cooling run used for prepoling. The red lines in Fig. 5(c) demonstrate that, even in this case, no canonical ferroelectric switching of P was achieved. These findings agree with the conclusions drawn from the dielectric experiments shown in Figure 4. However, the results shown in Fig. 5(c) at least indicate a certain impact of the external electrical field: The absolute values of ΔP [red line in the inset of Fig. 5(c)] reach significantly larger values than for the results obtained without applying a magnetic field during prepoling (blue line). Obviously, reversing the polarity of the prepoling field especially influences the polarization in the SkX and conical magnetic phases, resulting in a significant ΔP. Thus, while canonical ferroelectric switching behaviour is not observed, our measurements at least reveal a certain impact of external electrical fields on the polar state within the SkX phase.

Just as for the results in Fig. 3(a), from the minima in the P(H) curves of Fig. 5(c) the phase boundaries of the SkX phase were determined and included in the phase diagram [circles in Fig. 5(b)]. Some deviations compared to the results deduced from Fig. 3(a) are found, especially concerning the upper boundaries of the SkX and conical phases, which seem to extend to higher magnetic fields in the thinner sample 2 [cf. shaded areas in Fig. 5(b)]. Interestingly, for thin films of $Cu_2OSeO_3$ it was demonstrated[9] that the SkX phase is significantly extended. However, in the present case the slightly more extended SkX phase of sample 2 can be ascribed to demagnetization effects instead of a reduced dimensionality[13]. Using the relation of Aharoni[35], based on the geometry of the samples the demagnetization factors of samples 1 and 2 are estimated as $D_z = 0.43$ and $D_z = 0.74$, respectively. Indeed, after application of the appropriate correction factors for the measured magnetic fields of $(1 - D_z \chi)$, where $\chi$ is the magnetic susceptibility, the magnetic phase boundaries of both samples agree within few mT.

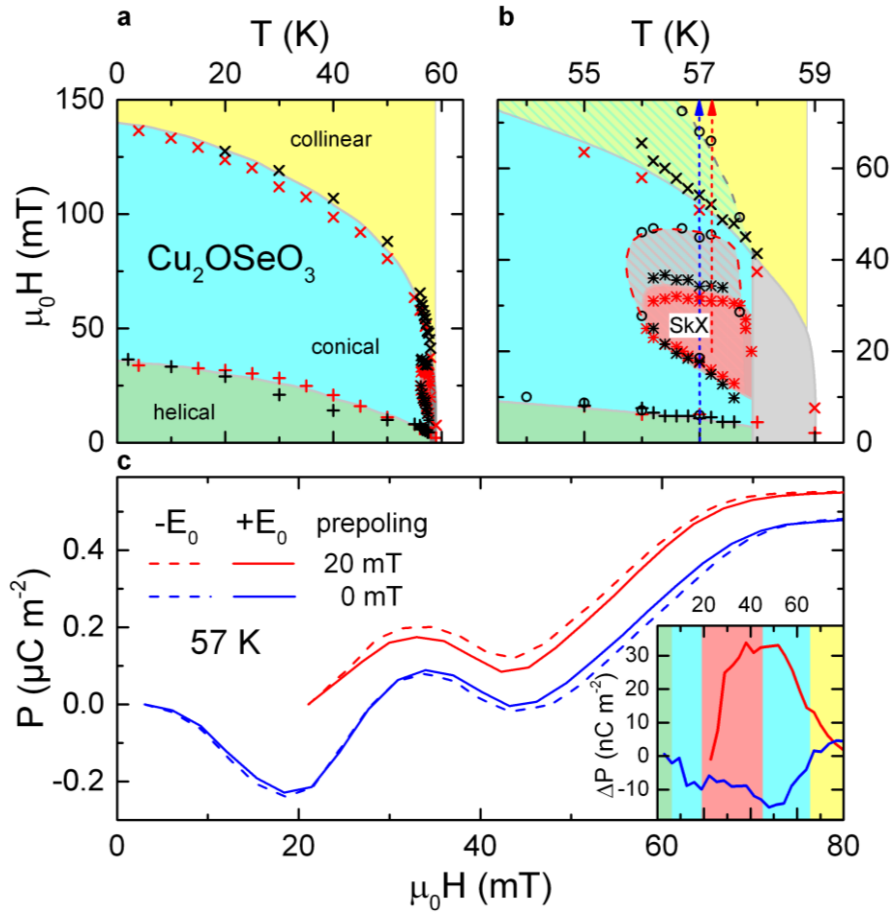

Fig 5. H,T-phase diagram obtained from measurements of two $Cu_2OSeO_3$ samples. The data points were deduced from magnetization (red symbols) and polarization measurements (black) for temperatures below $T_c$ (a) and in the vicinity of the SkX phase (b). The plusses and crosses mark the lower and upper phase boundaries of the conical phase, respectively (sample 1). The stars indicate the boundaries of the SkX-phase for sample 1. The results for sample 2 are denoted by circles. The colours indicate the helical (green), conical (blue), SkX (red) and collinear (yellow) magnetic phases as well as a regime of nearly critical spin fluctuations (grey). The hatched areas in (b) denote regions, where differences in the extension of the phases showed up for the two investigated samples. The red and blue arrows show the course of the magnetocurrent experiments leading to the P(H) curves presented in (c). (c) Magnetic-field dependent polarization derived from magnetocurrent measurements performed after cooling from temperatures above $T_c$ to 57 K with prepoling voltages of +330 kV/m (solid lines) and -330 kV/m (dashed) and with an applied external magnetic field of 20 mT (red) or 0 T (blue) (E ∥ H ∥ ⟨111⟩). (The experiments with 20 mT were performed at slightly higher temperature of 57.2 K.) The inset shows the differences between the polarizations obtained for positive and negative prepoling fields. The background colours in the inset denote the different phases of sample 2, used for these measurements (cf. frame (b)).

**Summary**

Bulk $Cu_2OSeO_3$ reveals a complex magnetic phase diagram including helical, conical and field-induced spin-collinear magnetic phases and, just below $T_c$, an additional state with reduced symmetry, the skyrmion phase. Skyrmions are particle-like objects of nanometre size, which possibly allow high data storage and low current manipulation in future data-storage devices. In this work, we have provided a detailed study of the magnetization, pyrocurrent, electric polarization and dielectric constants of this system as function of temperature and external magnetic fields, specifically focusing on the SkX phase. From the polarization and magnetization measurements, we have quantified the magnetoelectric coupling in all magnetic phases. A thorough analysis of the magnetic susceptibility and polarization obtained from magnetocurrent measurements, both at various temperatures and magnetic fields, allows the precise identification of phase boundaries.

The complexity of the magnetic phase diagram, the cancellation of ferroelectric polarization averaged over the domains of the multiple *q*-domain structure of the helical phase and the magnetoelectric effects, arising from the *d-p* hybridization, result in a complex magnetic-field dependence of the polarization and of the magnetoelectric coupling. It seems that all the electrical polarization in $Cu_2OSeO_3$ is only magnetic-field induced. We find no indications of spontaneous ferroelectric order in the temperature dependence of the dielectric constants and no indications for an electric-field induced switching of polarization. However, within the SkX phase the observed polarization at least to some extent can be influenced by an external electric field. It could be due to complex antiferroelectric order of the vortex-like-particles, thus leading to an unconventional multiferroic skyrmion. However, to clarify this complex mechanism, further investigations at higher electrical field are necessary.

**Methods**

High-quality single crystals of $Cu_2OSeO_3$ were grown by the standard chemical vapour phase method. Two crystals with different shapes were investigated: crystal 1 has an area of about A = 14 mm$^2$ and thickness of d = 2.1 mm. For the second crystal A = 5.4 mm$^2$ and d = 350 µm. Details about the crystal growth can be found in Ref.[36]. Magnetic properties were measured in the temperature range $2 \leq T \leq 300$ K using a SQUID magnetometer (Quantum Design MPMS-XL). For the isothermal magnetization measurements, $\mu_0H$ was varied between 0 and 150 mT. For measurements between 2 and 300 K and in external magnetic fields up to 9 T, a Quantum Design physical property measurement system was employed. To probe ferroelectric order, we measured the magnetoelectric current at fixed temperatures using a high-precision electrometer (Keithley 6517A). Before those magnetocurrent measurements, the samples were heated above $T_c$ followed by cooling with zero magnetic and electric field. For the poling experiment on crystal 2, during cooling external magnetic and electric fields were applied in ⟨111⟩ direction. The applied magnetic field sweep-rate was 10 Oe/s. The complex permittivity $\varepsilon^* = \varepsilon' - i\varepsilon''$ and the real part of the conductivity $\sigma'$ at frequencies $1\,Hz \leq \nu \leq 1\,MHz$ were determined using a frequency-response analyser (Novocontrol alpha-Analyser)[37]. For sample cooling between 10 and 300 K, a closed-cycle refrigerator was employed.

**Acknowledgements**

This work was supported by the Deutsche Forschungsgemeinschaft via the Transregional Collaborative Research Center TRR 80 (Augsburg/Munich/Stuttgart) and by the BMBF via ENREKON 03EK3015. We thank Dana Viewweg for performing the magnetic measurements and Andreas Bauer, Dennis Meier, Joachim Deisenhofer and Istvan Kezsmarki for useful discussions.

**Author contributions**

S.K. initiated the research. H.B. provided the samples for the experiments. S.K., A.L. and P.L. supervised the project. E.R. performed the magnetization, polarization and dielectric measurements. S.K. and A.L. wrote the paper with contributions from E.R. and P.L. All authors discussed the results and commented on the manuscript.

**Additional information**

Competing financial interests: The authors declare no competing financial interests.